\documentstyle[12pt]{article}
\newcommand{\be}[1]{\begin{equation}\label{#1}}
\newcommand{\ba}[1]{\begin{eqnarray}\label{#1}}
\newcommand{\ee}{\end{equation}}
\newcommand{\ea}{\end{eqnarray}}
\newcommand{\non}{\nonumber\\\rule{0pt}{30pt}}

\newcommand{\dis}{\displaystyle}
\newcommand{\eq}[1]{(\ref{#1})}
\newcommand{\sign}{\mathop{\rm sign}}
\newcommand{\Tr}{\mathop{\rm tr}}
\newcommand{\h}{\hat}

\newlength{\HFPP}       \HFPP5.4mm
\makeatletter
\@addtoreset{equation}{section}
\makeatother

\begin{document}

\begin{center}
{\large\bf Integral equations for the correlation functions of
the quantum one-dimensional Bose gas}

\vspace{2cm}

{\normalsize N. A. Slavnov}\\
{\it Steklov Mathematical Institute,\\
Gubkina 8, 117966,  Moscow, Russia\\}
nslavnov@mi.ras.ru

\end{center}

\vskip4pt

\begin{abstract}
\noindent
The large time and long distance behavior of the temperature
correlation functions of the quantum one-dimensional Bose gas
is considered. We  obtain integral equations, which solutions
describe the asymptotics.  These equations are closely related to the
thermodynamic Bethe Ansatz equations. In the low temperature limit
the solutions of these equations are given in terms of observables of
the model.
\end{abstract}

\section{Introduction}
This paper continues the series of works
\cite{KKS1}--\cite{IS}, devoted to the study of the correlation
functions of the quantum one-dimensional Bose gas with delta-function
interaction. Here we consider the problems related to the large time
and long distance asymptotics of the correlation functions. The main
result of this paper is a system of integral equations (see section
4), which solutions describe the asymptotics. The form of these
equations is close to the thermodynamic Bethe Ansatz equations, and
apparently their solutions are closely related to  observables of
the one-dimensional Bose gas model. Anyhow such the relationship does
exist in the low temperature limit.

The main object of our investigation is a temperature correlation
functions of local fields
\be{0tempcorrel}
\langle\Psi(0,0)\Psi^\dagger(x,t)\rangle_T=
\frac{\Tr\left( e^{-\frac HT}\Psi(0,0)\Psi^\dagger(x,t)\right)}
{\Tr e^{-\frac HT}}.
\ee
Here $T$ is a temperature, $H$ is the Hamiltonian
\be{0Hamilton}
{H}={\dis\int \,dx
\left({\partial_x}\Psi^{\dagger}(x)
{\partial_x} \Psi(x)+
c\Psi^{\dagger}(x)\Psi^{\dagger}(x)\Psi(x)\Psi(x)
-h \Psi^{\dagger}(x)\Psi(x)\right),}
\ee
where $c>0$ is the coupling constant, $h$---chemical potential. The
operators $\Psi(x,t)$ and $\Psi^{\dagger}(x,t)$ are the canonical Bose
fields
\be{0com}
[\Psi(x,t), \Psi^{\dagger}(y,t)]=\delta(x-y).
\ee
The equation of motion, corresponding to the Hamiltonian
\eq{0Hamilton} is refered as quantum Nonlinear Schr\"odinger equation
(quantum NLS).

For the reader's convenience we present here the list of the basic
equations for observables of the one-dimensional Bose gas at a finite
temperature. The energy of elementary excitation (particle)
satisfies the Yang--Yang equation \cite{YY}
\be{0YY}
\varepsilon(\lambda)=\lambda^2-h-\frac{T}{2\pi}
\int_{-\infty}^{\infty}K(\lambda,\mu)
\ln\left(1+e^{-\frac{\varepsilon(\mu)}{T}}
\right)d\mu,\qquad
K(\lambda,\mu)=\frac{2c}{c^2+(\lambda-\mu)^2}.
\ee
In the case of positive chemical potential the function
$\varepsilon(\lambda)$ has two real roots $\varepsilon(\pm q_T)=0$.
The total spectral density of the vacancies in the gas is given by
the equation
\be{0rhot}
2\pi\rho_t(\lambda)=1+\int_{-\infty}^{\infty}
K(\lambda,\mu)\vartheta(\mu)
\rho_t(\mu)\,d\mu,
\ee
where
\be{0Fermi}
\vartheta(\lambda)=\left(1+\exp\left[
\frac{\varepsilon(\lambda)}{T}\right]\right)^{-1}
\ee
plays the role of the Fermi weight.

The momentum of particle, being a function of the spectral parameter,
also can be found from the corresponding integral equation. However
below we shall need only the derivative of the momentum with respect
to the spectral parameter. This quantity coincides up to a coefficient
with the total density of the vacancies:
\be{0mement}
\frac{\partial k(\lambda)}{\partial\lambda}=
2\pi \rho_t(\lambda).
\ee
The velocity of the particle $v(\lambda)$ is
\be{0velocity}
v(\lambda)=
\frac{\partial \varepsilon}{\partial k}=
\frac{\varepsilon'(\lambda)}{k'(\lambda)}=
\frac{\varepsilon'(\lambda)}{2\pi \rho_t(\lambda)}.
\ee
Finally, we present also the integral equation for the scattering
phase
\be{0phase}
2\pi F(\lambda,\nu)=\int_{-\infty}^{\infty}
K(\lambda,\mu)\vartheta(\mu)
F(\mu,\nu)\,d\mu+i\ln\left(\frac{ic+\lambda-\nu}
{ic+\nu-\lambda}\right).
\ee
Here and further the branch of a logarithm is chosen such that
$-\pi<\arg\ln z\le\pi$. The quantity $S(\lambda,\nu)=\exp\{2\pi
iF(\lambda,\nu)\}$ is the scattering matrix of two particles
possessing momenta $k(\lambda)$ and $k(\nu)$.

All the thermodynamics observables listed here will be needed later
on for the description of the correlation function asymptotics.

We shall consider the behavior of the correlation function
\eq{0tempcorrel} at $x\to\infty$, $t\to\infty$ and fixed ratio
$\lambda_0=x/2t$. The one-dimensional Bose gas is massless model,
therefore  the correlation  function \eq{0tempcorrel}
decay as a power at zero temperature
$$
\langle\Psi(0,0)\Psi^\dagger(x,t)\rangle_0
\longrightarrow C t^{-\Delta_0}.
$$
The exponent $\Delta_0$  can be computed in the framework of the
conformal field theory \cite{BPZ}. At finite temperature the
asymptotics replaces by the exponential one
\be{0asy}
\langle\Psi(0,0)\Psi^\dagger(x,t)\rangle_T
\longrightarrow C t^{-\Delta}e^{-t/r}.
\ee
In other words the zero temperature turns out to be the point of
phase transition. The methods of the conformal field theory can be
applied at a finite temperature only approximately, for instance,
for calculations of the low temperature contributions into the
correlation radius \cite{BPZ}.

The determinant representation method, used in \cite{KKS1}--\cite{IS},
allows one in principle to consider correlation functions at
arbitrary temperature. This method is especially powerful for study
of their asymptotics. In the present paper we focus our attention at
the main, exponential law of decay in the equation \eq{0asy}. We
shall not concern neither the pre-exponential factor  $t^{-\Delta}$,
nor the constant coefficient $C$ in \eq{0asy}. In other words our
goal is to compute the correlation radius.

Let us give now the general content of the paper. In the next section
we present a brief description of the method, based on the
representation of the correlation functions in terms of Fredholm
determinants. Here we also discuss the questions related to the
construction of the dual fields. In particular we explain the cause
of these auxiliary quantum operators appearance and their role in the
calculation of the correlation functions. On the whole the section 2
is a sort of review. The reader can find the details in the papers
\cite{KKS1}--\cite{IS}, mentioned above. In the section 3 the methods
of the averaging of the operators, depending on dual fields, is
developed. Here we define the `averaging mapping', associating quantum
operators with certain classical functions. This mapping allows us to
compute the asymptotics of the correlation function
\eq{0tempcorrel}. This is done in the section 4. Here the system of
integral equations, describing the asymptotics, is given. In the last
section we consider some specific cases of the correlation function
\eq{0tempcorrel}. In particular we prove that in the low temperature
limit our results exactly coincide with the results of the paper
\cite{BPZ}.

\section{Dual fields}

The method of the Fredholm determinant representation for evaluation
of the correlation functions is described in details, for example,
in the book \cite{KBI}. At the first stage of this method one need to
represent a correlation function in terms of a determinant of an
integral operator, which kernel depends on distance and time
(and other physical parameters of the model). For the correlation
function \eq{0tempcorrel} such the representation was found in
\cite{KKS1}. Further investigation of the Fredholm determinant
obtained, can be performed via the methods of classical exactly
solvable equations. The matter is that the determinants mentioned
above turns out to be $\tau$-functions of classical integrable
systems. In particular, for the model of the quantum one-dimensional
Bose gas they can be expressed in terms of solutions of classical
forms NLS (in the simplest case of two-points  field correlation
function in the free fermionic limit the corresponding differential
equation is scalar Nonlinear Schr\"odinger equation). In turn the
solutions of classical integrable equations can be found by means of
the Riemann--Hilbert problem approach. In particular, for the
calculation of the large time and long distance asymptotics the
nonlinear steepest descent method is used \cite{DZ}, allowing in
principle to obtain the complete asymptotic expansion for
correlation functions. Thus, the computation of the correlation function
asymptotics in the framework of the method described, consists of two
stages:
\begin{enumerate}
\item
representation of the correlation function in the Fredholm
determinant form;
\item
computation of the asymptotics of the determinant obtained via the
methods of the Ri\-e\-mann--Hil\-bert problem and classical exactly
solvable equations.  \end{enumerate}

However that is how the matter stands only for the free fermionic
models (in the quantum NLS this corresponds to the limit of infinite
coupling constant:  $c\to\infty$). Out off the free fermionic point
the situation is more complicated, and one more step should be added to
the scheme described above:
\begin{enumerate}
\setcounter{enumi}{2}
\item
averaging with respect to an auxiliary vacuum.
\end{enumerate}
The present paper is devoted just to the last stage of calculations,
therefore we consider this problem more detailed.

In order to obtain the Fredholm determinant representation for the
correlation functions of non-free fermionic models (finite coupling
constant for one-dimensional Bose gas), one need to introduce
auxiliary quantum operators---Korepin's dual fields \cite{K}. This
necessity is caused by the existence of a non-trivial $S$-matrix.
Consider the main idea in brief. For evaluation of a correlation
function of an operator $\cal O$ with respect to a state $|a\rangle$
one can use the form factor decomposition
\be{2formdec}
\langle a|{\cal O}(x,t){\cal O}(0,0)|a\rangle=
\sum_{|b\rangle}\left|\langle a|{\cal O}(0,0)|b\rangle\right|^2
e^{D_{ab}(x,t)}.
\ee
Here the state $|b\rangle$ runs through a complete set, the function
$D_{ab}(x,t)$ describes the dispersion law of the model. The quantity
$\langle a|{\cal O}(0,0)|b\rangle$ is refered as a form factor of the
operator ${\cal O}$, and, for example, in the quantum NLS it can be
computed explicitly via the algebraic Bethe Ansatz. Thus, for
${\cal O}=\Psi$ the form factor is proportional to a certain
determinant:
\be{2ffNS}
\langle a|\Psi(0,0)|b\rangle\sim\det M.
\ee
We would like to emphasize that in distinction of the quantum field
theory models here the form factors are constructed in the bare Fock
vacuum, but not in the physical one. If the states $|a\rangle$ and
$|b\rangle$ are parameterized by the spectral parameters  $\{\lambda\}$
and $\{\mu\}$ respectively, then the entries $M_{jk}$ are functions
of these spectral parameters
$$
M_{jk}=M_{jk}(\{\lambda\},\{\mu\}).
$$
The type of dependency of $M_{jk}$ on the spectral parameters is
extremely important. In fact, in order to obtain a determinant
representation for the mean value \eq{2formdec}, we need to reduce
the infinite sum of determinants in the r.h.s. \eq{2formdec} to a single
determinant. This can be done only for very special form of
matrices $M$, namely, if $M_{jk}=M(\lambda_j,\mu_k)$, i. e. the
entries are parameterized by a single two-variable function
$M(x,y)$, taken in the points $x=\lambda_j$ and $y=\mu_k$. These are
so called local matrices. In the reality the entries of the matrix
$M_{jk}$, describing the form factor \eq{2ffNS}, contain the products
of two-particle $S$-matrices
\be{2Smatr}
S(\lambda,\mu)=-\frac{ic+\lambda-\mu}{ic+\mu-\lambda}.
\ee
with respect to all parameters $\{\lambda\}$  and $\{\mu\}$.
This leads to the effect that for interacting fermions the entries
$M_{jk}$  depend on all spectral parameters
$M_{jk}=M(\{\lambda\},\{\mu\})$ (non-local matrices). As a result, it
becomes impossible to reduce the sum of determinants \eq{2formdec} to
a single determinant. On the other hand, in the free fermionic limit
$(c\to\infty)$ the $S$-matrix becomes trivial, and, as a consequence,
the entries $M_{jk}$ depend only on two variables
$M_{jk}=M(\lambda_j,\mu_k)$.  The details and all explicit equations
can be found in \cite{KKS1}.

The introducing of dual fields updates the situation. In order to
reduce the matrix $M$ to the local form it is sufficient to factorize
the $S$-matrix
\be{2factor}
S(\lambda,\mu)=f_1(\lambda)f_2(\mu),
\ee
where $f_j$ are some functions. Of course, such a factorization is
impossible for the $S$-matrix \eq{2Smatr}, if $f_j$ are classical
functions. However one can easily achieve this by means of quantum
objects.

Consider two creation operators $q_\psi(\lambda)$ and
$q_\phi(\lambda)$, and two annihilation operators $p_\psi(\lambda)$,
$p_\phi(\lambda)$, acting in an auxiliary Fock space as
\be{2action}
(0|q_\psi(\lambda)=(0|q_\phi(\lambda)=0,\qquad
p_\psi(\lambda)|0)=p_\phi(\lambda)|0)=0.
\ee
The non-vanishing commutation relations are
\be{2commut}
{}[p_\phi(\lambda),q_\psi(\mu)]=-[p_\psi(\lambda),q_\phi(\mu)]=
\ln\left( \frac{ic+\lambda-\mu}
{ic+\mu-\lambda}\right).
\ee
The dual fields are linear combinations
\be{2dualfield}
\psi(\lambda)=q_\psi(\lambda)+p_\psi(\lambda),\qquad
\phi(\lambda)=q_\phi(\lambda)+p_\phi(\lambda).
\ee
Obviously, due to \eq{2commut} the dual fields commute with each
other
\be{2Abel}
{}[\psi(\lambda),\psi(\mu)]=[\phi(\lambda),\phi(\mu)]=
[\psi(\lambda),\phi(\mu)]=0.
\ee
On the other hand, the vacuum expectation value of the expressions,
containing the dual fields, might be non-trivial
\be{2vacexp}
S(\lambda,\mu)=-\frac{ic+\lambda-\mu}{ic+\mu-\lambda}=
-(0|e^{\phi(\lambda)}e^{\psi(\mu)}|0).
\ee
Thus, the $S$-matrix can be presented as vacuum expectation value of
the factorized expression of the form \eq{2factor}. Due to the
property \eq{2Abel}, the determinants, depending on the dual fields,
are well defined. As a result the form factors can be presented in the
form
\be{2vacexpff}
\langle a|{\cal O}(0,0)|b\rangle\sim\det M=
(0|\det\tilde M(\phi,\psi)|0),
\ee
where matrix $\tilde M$ is now local one, i. e. $\tilde
M_{jk}=\tilde M(\lambda_j,\mu_k)$.

Similar representation can be obtained for the square of the absolute
value of the form factor \eq{2formdec}, and the summation with respect
to complete set $\{|b\rangle\}$ can be performed under the symbol
of the vacuum expectation value. Eventually  we arrive at the
representation for the mean value \eq{2formdec} in terms of
the vacuum expectation value in the auxiliary Fock space of a
determinant of a certain matrix, depending on the dual fields. In the
thermodynamic limit this matrix turns into an integral operator, and
we obtain
\be{2vacexpcf}
\langle{\cal O}(x,t){\cal O}(0,0)\rangle\sim (0|\det(I+V)|0).
\ee
Here the kernel of the operator $V$ depends on the dual fields
$\phi(\lambda)$ and $\psi(\lambda)$, however due to
\eq{2Abel}, the Fredholm determinant is well defined object.

We would like to emphasize that here we have described only the
main idea of the dual fields method. In the specific cases some
modifications are possible, in particular, sometimes one need to
introduce additional dual fields. The reader can find the detailed
formul\ae\ in \cite{KKS1}.  However the basic idea of the method always
consists of the applying of factorization \eq{2factor}, which allows
to reduce the determinants of non-local matrices to the local ones.

Thus the correlation function of the quantum one-dimensional Bose
gas turns out to be proportional to the vacuum expectation value of
the Fredholm determinant, functionally depending on the dual fields.
Nevertheless the asymptotic analysis of this determinant can be
performed in the same manner, as for free fermionic limit. Due to
the commutation relations \eq{2Abel} at the certain stage of the
calculations one can consider the  dual fields as some classical
functions, which are analytical in the strip $|\Im\lambda|<c/2$. The
last property follows from the representation of the dual fields in
terms of the  canonical Bose fields
\be{2repfields}
\begin{array}{l}
{\dis
\phi(\lambda)=\frac{1}{\sqrt{2\pi}}\int\ln\left(\frac
{\frac{ic}{2}+\nu-\lambda}{\frac{ic}{2}-\nu+\lambda}\right)
(\varphi_1^\dagger(\nu)+\varphi_1(\nu))\,d\nu;}\non
{\dis
\psi(\lambda)=\frac{c}{\sqrt{2\pi}}\int\frac
{\varphi_2^\dagger(\nu)+\varphi_2(\nu)}{(\nu-\lambda)^2+\frac{c^2}{4}}
\,d\nu,}
\end{array}
\ee
where
\be{cancom}
{}[\varphi_1(\lambda),\varphi_2^\dagger(\mu)]=
{}[\varphi_2(\lambda),\varphi_1^\dagger(\mu)]=\delta(\lambda-\mu),
\qquad
\varphi_j(\lambda)|0)=0,\qquad (0|\varphi_j^\dagger(\lambda)=0.
\ee
It is easy to check, that commutation relations between the operators
$p_{\psi,\phi}$ and $q_{\psi,\phi}$, defined by
\eq{2repfields}, exactly coincide with \eq{2commut}.

Via the Riemann--Hilbert problem and classical integrable
equations methods one can find the large time and long distance
behavior of the Fredholm determinant \eq{2vacexpcf}  (see
\cite{KKS2}-\cite{IS}).  The asymptotic expression obtained is a
functional of the dual fields, which up to this moment are considered
as some {\it classical} functional parameters. In order to calculate
the asymptotics of the correlation function one need now to remember
about quantum nature of these operators and to average the
obtained expression with respect to auxiliary vacuum.

Here, however, one important moment exists, which had been discussed in
details in \cite{IS}. The matter is, that the procedures of averaging
and calculation of the asymptotics, strictly speaking, do not
commute with each other. The asymptotics of the correlation function
is equal to the asymptotics of the vacuum expectation value
\eq{2vacexpcf}, but not to the vacuum expectation value of the
asymptotics. Just because of this reason the results of the work
\cite{KS2}, where some questions of averaging of the expressions
containing dual fields, were considered, can be applied to the
evaluation of the correlation functions only in some measure.
Nevertheless it was proved in the mentioned above paper \cite{IS}
that there exist asymptotic representations, which are stable with
respect to the procedure of averaging, i. e.
$$ \mbox{asymptotics(vacuum expectaion value)}=
\mbox{vacuum expectation value(asymptotics)}.
$$
Such the representation for the Fredholm determinant was obtained in
\cite{IS}, and further we shall deal just with this one.


\section{Averaging mapping}

The asymptotic formul\ae, obtained in \cite{IS} for the Fredholm
determinant in terms of the dual fields, are rather bulky
(see \eq{4main1}, \eq{4main2} below). Therefore before averaging
of these expressions we consider several more simple, but at the
same time quite general examples. The technique, developed in the
present section, will allow us to average the equations
\eq{4main1}, \eq{4main2} easily.

Consider two dual fields $\phi$ and $\psi$, defining by
\eq{2dualfield}. Let the commutation relations between the creation
and annihilation operators are
\be{3commut1}
{}[p_\psi(\lambda),q_\phi(\mu)]=\xi(\lambda,\mu),\qquad
[p_\phi(\lambda),q_\psi(\mu)]=\xi(\mu,\lambda),
\ee
where $\xi(\lambda,\mu)$ is a two-variable function. Here we do not
need the explicit form of this function, however it is easy to see that
putting $\xi(\lambda,\mu)=\ln(ic+\mu-\lambda)-\ln(ic+\lambda-\mu)$, we
reproduce the commutation relations \eq{2commut}. Besides the relations
\eq{3commut1} we introduce an additional one
\be{3commut2}
{}[p_\psi(\lambda),q_\psi(\mu)]=\eta(\lambda,\mu)=
\eta(\mu,\lambda),
\ee
where $\eta(\lambda,\mu)$ is a symmetric two-variable function. Below
we shall see that the additional relation \eq{3commut2} does not affect
essentially on the final result, however for some time we do not put the
function $\eta(\lambda,\mu)$ equal to zero.

It is easy to see that the main property of the dual
fields \eq{2Abel} is still valid. However the vacuum
expectation values may be non-trivial as before.

The main goal of this section is evaluation of the mean value of
the following form:
\be{3mainaim}
(0|e^{{\cal F}(\psi,\phi)}F(\phi)|0).
\ee
Here ${\cal F}(\psi,\phi)$ and $F(\phi)$ are operator-valued functionals
(later on functionals for brief) of the dual fields. Functional
$F$ does not depend on the field $\psi$, while $\cal F$ depends
on this field linearly
\be{3linear}
\frac{\delta^2{\cal F}}{\delta\psi^2}=0.
\ee
All the functionals (and functions) of the dual fields should be
understood in the sense of formal series.

We are starting with the formula, obtained in \cite{KS2}
\be{3vacexp}
(0\vert \exp\left\{ \sum_{k=1}^N\alpha \psi(\lambda_k)
 f_k\bigl(\phi(\mu_{k})\bigr) \right\}
F\Bigl(\phi(\nu)\Bigr) \vert 0)
=F\biggl(\sum_{m=1}^{N}z_m\xi(\lambda_m,\nu)\biggr)
\frac{E(\eta)}{\det g},
\ee
where
\be{3E}
E(\eta)=\exp\left\{\frac{1}{2}\sum_{n,m=1}^{N}
z_nz_m\eta(\lambda_n,\lambda_m)\right\}.
\ee
Here in the l.h.s. $F\Bigl(\phi(\nu)\Bigr)$ is a functional (or a
function) of the field $\phi$, $\alpha$ is a complex parameter, functions
$f_k(z)$ assumed to be holomorphic in a vicinity of the origin. In the
r.h.s. the quantities $z_j$ are the roots of the system
\be{3system}
z_j - \alpha f_j\biggl( \sum_{m=1}^N z_m
\xi(\lambda_m,\mu_{j})\biggr)=0.
\ee
If the system has several solutions, then one should choose the single
one, approaching zero at  $\alpha\to0$  (see \cite{KS2}). Finally,
in the denominator of \eq{3vacexp} the Jacobian of the system
\eq{3system} is placed
\be{3matrix}
g_{jk} = \frac{\partial}{\partial z_k}
\left[z_j - \alpha f_j\biggl( \sum_{m=1}^N z_m
\xi(\lambda_m,\mu_{j})\biggr)\right].
\ee

It is suitable to modify the result \eq{3vacexp}. Namely, let us
introduce a function $w(u)$
\be{3wu}
w(u)=\sum_{m=1}^{N}z_m\xi(\lambda_m,u).
\ee
Multiplying each of the equations \eq{3system}
by  $\xi(\lambda_j,u)$ and summing up all of them, we obtain
\be{3equation}
G(w)\equiv w(u)-\alpha\sum_{m=1}^{N}f_m\biggl(w(\mu_{m})\biggr)
\xi(\lambda_m,u)=0.
\ee
Thus instead of the system \eq{3system} we obtain only one equation
\eq{3equation} for the function $w(u)$, which has to be valid, however,
for any value of $u$. The roots of the system $z_j$ can be expressed
in terms of $w(u)$ as follows
\be{3roots}
z_j =\alpha f_j\bigl(w(\mu_{j})\bigr).
\ee
It is easy to see also that
\be{3Jakob}
\det g_{jk}=\det\left[\frac{\delta G(w(u))}{\delta w(u')}\right].
\ee
Thus the result \eq{3vacexp} turns into
\be{3result}
(0\vert \exp\left\{ \sum_{k=1}^N\alpha \psi(\lambda_k)
 f_k(\phi(\mu_{k})) \right\}
F\Bigl(\phi(\nu)\Bigr) \vert 0)
=F\Bigl(w(\nu)\Bigr)\frac{E(\eta)}
{\det \frac{\delta G}{\delta w}}.
\ee

Let now the functional $F(\phi)$ is simultaneously a function of
some parameter $t$: $F(\phi)=F(\phi|t)$. This parameter  will
play the role of time in the asymptotic formul\ae\ for the correlation
function. Then the equation \eq{3result} can be written in the form
\be{3result1}
(0\vert \exp\left\{ \sum_{k=1}^N\alpha \psi(\lambda_k)
 f_k(\phi(\mu_{k})) \right\}
F\Bigl(\phi(\nu)|t\Bigr) \vert 0)
=C F\Bigl(w(\nu)|t\Bigr),
\ee
where $C$ is a constant, does not depending on $t$. Constant factors
of the such type will be not interesting for us in computation of the
correlation function asymptotics. Just because of this reason on can
put the function $\eta(\lambda,\mu)$  equal to zero, since it
contributes only into the common constant multiplier and does not
affect on the dependency of the r.h.s. of \eq{3result1} on $t$.

Thus we see that under the averaging of the expressions of the form
\eq{3vacexp} a mapping of the operator $\phi$ into the classical
function $w$ takes place
\be{3mapaver}
\phi(\nu)\to w(\nu),\qquad
F\Bigl(\phi(\nu)\Bigr) \to F\Bigl(w(\nu)\Bigr),
\ee
where $w(\nu)$ is the solution of the equation \eq{3equation}.
We call the mapping \eq{3mapaver} (as well as more general equations
\eq{3genres}, \eq{3geneq})  averaging mapping.

Consider now a generalization of the functional
\be{3general}
{\cal F}(\psi,\phi)=\alpha\sum_{k=1}^{N}\psi(\lambda_k)
f_k\Bigl(\phi(\mu_k)\Bigr),
\ee
standing in the exponent of \eq{3result1}. It is clear, that one
can easily include the derivatives of the fields
$\psi^{(m)}(\lambda)$ and $\phi^{(m)}(\lambda)$ into this functional.
One can consider also the continuous limit, when the sum with respect to
$k$ turns into an integral (this was done in \cite{KS2}). Besides it is
possible to consider infinite series instead of finite sum with
respect to $k$. In all listed cases the result of averaging can be
easily obtained by the corresponding limiting procedure in the
equation \eq{3result1} (or in \eq{3result}, if more precise results
are needed).  It is important only that  ${\cal F}(\psi,\phi)$
would be always linear with respect to the field $\psi$ and its
derivatives, while the dependency of $\cal F$ and $F$ on the field
$\phi$ might be rather complicated.

Thus we arrive at the following formula
\be{3genres}
(0|e^{{\cal F}(\psi(\cdot),\phi)}F\Bigl(\phi(\nu)|t\Bigr)|0)=
C F\Bigl(w(\nu)|t\Bigr),
\ee
where the function $w(u)$ satisfies the following equation
\be{3geneq}
w(u)-{\cal F}\Bigl(\xi(\cdot,u),w\Bigr)=0.
\ee

As for functionals, depending on the field $\psi$ in non-linear
way, such a simple formula, generally speaking, does not exist.
However in scattered instances it is possible to generalize the
averaging mapping for more complicated functionals. Below we consider
an example of functionals, which are `asymptotically linear' with respect
to the field $\psi(\lambda)$.

Consider a mean value
\be{3expLambda}
(0|e^{{\cal F}(\psi(\cdot),\phi,\h\Lambda)}
F_c\Bigl(\phi(\nu),\h\Lambda\Bigr)
e^{tF_e\bigl(\phi(\nu),\h\Lambda\bigr)}|0),
\ee
We are interesting of the behavior of this mean value at $t\to\infty$.
Let the explicit dependency of the functional
${\cal F}(\psi(\cdot),\phi,\h\Lambda)$ on the field $\psi$ is linear
as before. Besides let $\cal F$ and $F$ depend on  $\psi$
implicitly through an operator $\h\Lambda$, which is defined as the
`root' of the equation
\be{3defLambda}
\h\Lambda=\lambda_0+\frac{i}{2t}\psi'(\h\Lambda),
\qquad \h\Lambda\to\lambda_0\quad\mbox{at}\quad
t\to\infty.
\ee
Here $\lambda_0$ is a constant, prime means the derivative with respect
to the argument. The definition \eq{3defLambda} is formal, and one
should understand it in the sense of a series with respect to
$t^{-1}$. If we replace in \eq{3defLambda} the operator $\psi$ by a
complex function holomorphic in a vicinity of $\lambda_0$, then for
arbitrary function $f(z)$, holomorphic in the same vicinity, we have
under the corresponding choice of the contour
\be{3Cauchi}
f(\h\Lambda)=\frac{1}{2\pi i}\oint\,dz f(z)
\frac{1-\frac{i}{2t}\psi''(z)}
{z-\lambda_0-\frac{i}{2t}\psi'(z)}.
\ee
For large enough $t$ one can expand the denominator in \eq{3Cauchi}
into the absolutely convergent series, what gives
\be{3series2}
f(\Lambda)=\sum_{n=0}^{\infty}\frac1{n!}
\left(\frac{i}{2t}\right)^n\frac{d^{n}}
{d\lambda_0^{n}}
\left(1-\frac{i}{2t}\psi''(\lambda_0)\right)
\left[(\psi'(\lambda_0))^nf(\lambda_0)\right].
\ee
In particular
\be{3series1}
\Lambda=\lambda_0+
\sum_{n=1}^{\infty}\frac1{n!}
\left(\frac{i}{2t}\right)^n\frac{d^{n-1}}
{d\lambda_0^{n-1}}\left[(\psi'(\lambda_0))^n\right].
\ee
We accept the formula \eq{3series1} as the definition of the
operator $\h\Lambda$. Accordingly functions (and functionals) of
this operator should be understood in the sense
\eq{3series2}. Obviously, for large value of $t$ the difference between
$\lambda_0$ and $\h\Lambda$ behaves as $t^{-1}\to0$. Therefore
we call the functional ${\cal F}(\psi(\cdot),\phi,\h\Lambda)$
asymptotically linear with respect to $\psi$. Nevertheless the
replacement $\h\Lambda$ by $\lambda_0$ in \eq{3expLambda} is illegal,
since $\psi$ is unbounded operator.

Using \eq{3series1} one can rewrite the mean value \eq{3expLambda}
in the form
$$
(0|\sum_{n=0}^{\infty}\frac1{n!}
\left(\frac{i}{2t}\right)^n\frac{\partial^{2n}}
{\partial\lambda_0^{n}\partial\beta^n}
\left[e^{\beta\psi'(\lambda_0)
+{\cal F}(\psi(\cdot),\phi,\lambda_0)}
\left(1-\frac{i\psi''(\lambda_0)}{2t}\right)
F_c\Bigl(\phi(\nu),\lambda_0\Bigr)
e^{tF_e\bigl(\phi(\nu),\lambda_0\bigr)}|0)\right]_{\beta=0}
$$
what in turn reduces to the double integral
\ba{3abc}
&&{\dis\hspace{-2cm}
-\frac{2t}{\pi}(0|\int_0^\infty\rho\,d\rho
e^{2i(t+i0)\rho^2}\oint\frac{dz}{z}
e^{\rho z^{-1} \psi'(\lambda_0+\rho z)
+{\cal F}(\psi(\cdot),\phi,\lambda_0+\rho z)}
}\non
&&{\dis\hspace{2cm}
\times\left(1-\frac{i}{2t}\psi''(\lambda_0+\rho z)\right)
F_c\Bigl(\phi(\nu),\lambda_0+\rho z\Bigr)
e^{tF_e\bigl(\phi(\nu),\lambda_0+\rho z\bigr)}|0),}\nonumber
\ea
where the integral over  $z$ is taken with respect to a small
contour around the origin. Recall that we are interesting in the
asymptotic behavior of this mean value at $t\to\infty$. It is not
difficult to check that in this case the term, containing the second
derivative  $\psi''$, contributes only into a constant factor and
hence, it can be removed. Then the integrand takes the form considered
above, and we can average it. We have
\be{3averag}
-\frac{2t}{\pi}\int_0^\infty\rho\,d\rho
e^{2i(t+i0)\rho^2}\oint\frac{dz}{z}
C(\rho,z,w)e^{tF_e\bigl(w(\nu),\lambda_0+\rho z\bigr)},
\ee
where
\be{3eqinter}
w(u,\rho,z)-\rho z^{-1} \xi'(\lambda_0+\rho z,u)-
{\cal F}(\xi(\cdot,u),w,\lambda_0+\rho z)=0.
\ee
It remains now to take the integral in \eq{3averag}. For $t\to\infty$
it can be estimated via the steepest descent method, and we arrive at
\be{3expLambdares}
(0|e^{{\cal F}(\psi(\cdot),\phi,\h\Lambda)}
F_c\Bigl(\phi(\nu)\Bigr)e^{tF_e\bigl(\phi(\nu)\bigr)}|0)
\to Ce^{2it(\Lambda-\lambda_0)\beta
+tF_e\bigl(w(\nu,\Lambda,\beta),\Lambda\bigr)}.
\ee
Here the function $w(u)$ satisfies the equation
\be{3eqwLb}
w(u,\Lambda,\beta)-\beta\xi'(\Lambda,u)-
{\cal F}(\xi(\cdot,u),w,\Lambda)=0,
\ee
where numbers $\Lambda$ and $\beta$ are defined by the equations
of the saddle point:
\be{3saddle}
\begin{array}{l}
{\dis 2i\beta+\frac{\partial}
{\partial\Lambda}F_e(w,\Lambda)=0,}\non
{\dis 2i(\Lambda-\lambda_0)+
\frac{\partial}{\partial\beta}F_e(w,\Lambda)=0.}
\end{array}
\ee
Thus, in the example considered we again deal with the averaging
mapping, however of more general form
\be{3genaver}
\phi(\nu)\to w(\nu),\qquad
\h\Lambda\to\Lambda,\qquad
F(\phi,\h\Lambda)\to F(w,\Lambda).
\ee
%

\section{Asymptotics of the correlation function}

Now we are ready to consider the asymptotics of the correlation
function \eq{0tempcorrel}. There are two different phases in the model
of the one-dimensional Bose gas, corresponding to the negative and
positive chemical potential $h$ in the Hamiltonian \eq{0Hamilton}.
While at $h<0$ the ground state coincides with the bare Fock vacuum,
at $h>0$ the ground state is the Fermi sphere. At a finite temperature
the difference is not so significant, but nevertheless the sign of
the chemical potential affects on the asymptotics of the correlation
function. The corresponding formul\ae\ were obtained in the paper
\cite{IS}.  Let us present here the explicit expressions.
\ba{4main1}
&&\hspace{-13mm}
{\dis\langle\Psi(0,0)\Psi^\dagger(x,t)\rangle_T\longrightarrow
(0|C_{-}(\phi,\tilde\phi|\lambda_0)
(2t)^{-\frac{(\nu(\h\Lambda)+1)^2}{2}}
e^{\psi(\h\Lambda)+it{\h\Lambda}^{2}-ix\h\Lambda-iht}}\non
&&{\dis\hspace{-4mm}\times\exp\left\{
\frac{1}{2\pi}\int_{-\infty}^\infty\biggl(x-2\lambda
t+i\psi'(\lambda)\biggr)\sign(\h\Lambda-\lambda)
\ln\left[\frac{e^{\frac{\varepsilon(\lambda)}{T}}-
e^{\phi(\lambda)\sign(\lambda-\h\Lambda)}}
{e^{\frac{\varepsilon(\lambda)}{T}}+1}\right]
\,d\lambda\right\}|0),}
\ea
for $h<0$, and
\ba{4main2}
&&\hspace{-13mm}
{\dis\langle\Psi(0,0)\Psi^\dagger(x,t)\rangle_T\longrightarrow
(0|C_{+}(\phi,\tilde\phi|\lambda_0)
(2t)^{-\frac{\nu^{2}(\h\Lambda)}{2}}
e^{\frac{1}{2}\psi(\h\Lambda_{1})+\frac{1}{2}\psi(\h\Lambda_{2})
+\frac{it}2({\h\Lambda_{1}}^{2}+{\h\Lambda_{2}}^{2})
-\frac{ix}2(\h\Lambda_{1}+\h\Lambda_{2})-iht}}\non
&&{\dis\hspace{-4mm}\times\exp\left\{
\frac{1}{2\pi}\int_{-\infty}^\infty\biggl(x-2\lambda
t+i\psi'(\lambda)\biggr)\sign(\h\Lambda-\lambda)
\ln\left[\frac{e^{\frac{\varepsilon(\lambda)}{T}}-
e^{\phi(\lambda)\sign(\lambda-\h\Lambda)}}
{e^{\frac{\varepsilon(\lambda)}{T}}+1}\omega(\lambda)\right]
\,d\lambda\right\}|0),}
\ea
for $h>0$.

Let us make necessary explanations. The equations  \eq{4main1},
\eq{4main2} were obtained under the assumption that $t\to\infty$,
$x\to\infty$, while the ratio  $x/2t=\lambda_{0}$ remains fixed.
Function $\varepsilon(\lambda)$ is the solution of the Yang--Yang
equation \eq{0YY}.

The most important objects for us are three dual fields
$\psi(\lambda)$, $\phi(\lambda)$ and $\tilde\phi(\lambda)$. The
commutation relations between corresponding creation and annihilation
operators are
\be{4commut}
\begin{array}{l}
{\dis
{}[p_\psi(\lambda),q_\phi(\mu)]=-[p_\phi(\lambda),q_\psi(\mu)]=
\xi(\lambda,\mu)\equiv
\ln\left( \frac{ic+\mu-\lambda}{ic+\lambda-\mu}\right),}\non
{\dis
{}[p_\psi(\lambda),q_\psi(\mu)]=[\tilde p_\phi(\lambda),q_\psi(\mu)]=
[p_\psi(\lambda),\tilde q_\phi(\mu)]=\eta(\lambda,\mu)
\equiv\ln\left(\frac{c^2}{(\lambda-\mu)^2+c^2}\right).}
\end{array}
\ee
As usual all the dual fields commute with each other. Recall that
just this property allows one at the certain stage of calculations to
consider these objects as classical functions. The equations
\eq{4main1}, \eq{4main2} were obtained in \cite{IS} just under such
treatment of the dual fields, although some operator features were
taken into account.  In particular, a special attention were paid to
the stability of the asymptotic formul\ae\ under the averaging
procedure with respect to the auxiliary vacuum, i. e. the corrections,
depending on the dual fields, would remain small after their
averaging. The presence in the asymptotic formul\ae\ of the value
$\h\Lambda$, which we dealt already in the previous section,
\be{4defLambda}
\h\Lambda=\lambda_0+\frac{i}{2t}\psi'(\h\Lambda),
\ee
is caused just by this reason.
All the expressions, containing the operator $\h\Lambda$, should be
understood in the sense of the series \eq{3series2}, \eq{3series1},
in particular, the integrals, depending on the sign-function, are
equal by definition
\be{4defsign}
\int_{-\infty}^{\infty}f\bigl(\lambda,\sign(\h\Lambda-
\lambda)\bigr)\,d\lambda=
\int_{-\infty}^{\h\Lambda}f(\lambda,1)\,d\lambda+
\int_{\h\Lambda}^{\infty}f(\lambda,-1)\,d\lambda.
\ee
One can use \eq{3series2} for each of the integrals in the r.h.s. of
\eq{4defsign}. In fact one can say that $\h\Lambda$ arises as the
result of partial summation of the asymptotic series for the
correlation function.

For operators $\h\Lambda_1$ and $\h\Lambda_2$ the situation is quite
analogous. If we consider dual fields as classical functions, then
$\h\Lambda_i$ are the roots of the equation
\be{4opereq}
\varepsilon(\h\Lambda_i)=T\sign(\h\Lambda-\h\Lambda_i)
\phi(\h\Lambda_i),\qquad
\h\Lambda_1\to-q_T,\quad\h\Lambda_2\to q_T,
\qquad\mbox{at}\qquad T\to0,
\ee
where $\pm q_T$---zeros of the Yang--Yang function:
$\varepsilon(\pm q_T)=0$. In other words the numbers $\h\Lambda_i$
correspond to the zeros of the expression, standing under the symbol
of the logarithm in the integral \eq{4main2}. In the real situation,
when  $\phi(\lambda)$ is the dual field,  $\h\Lambda_i$ become operators,
defining by the series similar to \eq{3series1}. We shall simply
demand the operator equation \eq{4opereq} to transform to the usual
equation under the averaging mapping
\be{4classeq}
\varepsilon(\Lambda_i)=T\sign(\Lambda-\Lambda_i)
w(\Lambda_i),\qquad
\Lambda_1\to-q_T,\quad\Lambda_2\to q_T,
\qquad\mbox{at}\qquad T\to0,
\ee

Function $\omega(\lambda)$, entering \eq{4main2}, is equal
$$
\omega(\lambda)=\sign(\lambda-\h\Lambda_1)
\sign(\lambda-\h\Lambda_2).
$$
These sign-functions are defined in the same manner as the
sign-functions of $\h\Lambda$ \eq{4defsign}.

The equations \eq{4main1}, \eq{4main2} contain also a function
$\nu(\h\Lambda)$, defining the power law of $t$
\be{4nu}
\nu(\h\Lambda)=\frac{1}{2\pi}\ln\left[\left(
\frac{e^{\frac{\varepsilon(\h\Lambda)}{T}}-e^{-\phi(\h\Lambda)}}
{e^{\frac{\varepsilon(\h\Lambda)}{T}}+1}\right)\left(
\frac{e^{\frac{\varepsilon(\h\Lambda)}{T}}-e^{\phi(\h\Lambda)}}
{e^{\frac{\varepsilon(\h\Lambda)}{T}}+1}\right)\right].
\ee

Finally the explicit form of factors $C_\pm(\phi,\tilde\phi|\lambda_0)$
is unknown (see \cite{IS}).  It is known, however, that they do not
depend on the field $\psi$, and they depend on the distance $x$ and
time $t$ only through the ratio $x/2t=\lambda_0$, which, recall,
remains fixed.

In spite of the asymptotic formul\ae\  \eq{4main1},
\eq{4main2} look rather complicated, nevertheless they belong to the
class of functionals of the dual fields, considered in the previous
section. They are asymptotically linear with respect to the field
$\psi(\lambda)$.  The operators $\h\Lambda_i$ are some
(implicit) functionals of the field $\phi(\lambda)$.
The presence of the third field  $\tilde\phi(\lambda)$, as well as
existence of unknown factors $C_\pm$, should not worry us, since due
to the trivial commutation relations
$$
{}[\tilde
p_\phi(\lambda),q_\phi(\mu)]= [p_\phi(\lambda),\tilde q_\phi(\mu)]=0
$$
one can easily see, that the averaging of the field $\tilde\phi$ may
contribute only into the common constant factor. Since our goal is to
compute the correlation radius, this factor is not essential, and we
even can put $\tilde\phi=0$.

Thus, applying the averaging mapping \eq{3genaver} to the equations
\eq{4main1}, \eq{4main2} and neglecting the common constant factor
and power law dependency on $t$, we have
\be{4asympt}
\langle\Psi(0,0)\Psi^\dagger(x,t)\rangle_T\longrightarrow
e^{-t/r_\pm},
\ee
where $r_\pm$ correspond to the positive and negative chemical
potential respectively. We shall call $r_\pm$ correlation radii,
although, strictly speaking, they are not necessary real.

For $h<0$
\be{4radneg}
-t/r_-=
2it(\Lambda-\lambda_0)\beta+it{\Lambda}^{2}-ix\Lambda-iht
+\frac{1}{2\pi}\int_{-\infty}^\infty(x-2\lambda t)
\sign(\Lambda-\lambda)
\ln\left[\frac{e^{\frac{\varepsilon(\lambda)}{T}}-
e^{w(\lambda)\sign(\lambda-\Lambda)}}
{e^{\frac{\varepsilon(\lambda)}{T}}+1}\right]
\,d\lambda.
\ee
Here function $w(u)$, depending on parameters $\Lambda$ and
$\beta$, is the solution of the integral equation
\be{4inteqneg}
w(u)=\xi(\Lambda,u)+i\beta K(\Lambda,u)-
\frac{1}{2\pi}\int_{-\infty}^\infty K(u,\lambda)
\sign(\Lambda-\lambda)
\ln\left[\frac{e^{\frac{\varepsilon(\lambda)}{T}}-
e^{w(\lambda)\sign(\lambda-\Lambda)}}
{e^{\frac{\varepsilon(\lambda)}{T}}+1}\right]
\,d\lambda,
\ee
where
\be{4xiK}
\xi(\lambda,\mu)=
\ln\left( \frac{ic+\lambda-\mu}{ic+\mu-\lambda}\right),
\qquad
K(\lambda,\mu)=-i\frac{\partial}{\partial\lambda}\xi(\lambda,\mu)=
\frac{2c}{(\lambda-\mu)^2+c^2}.
\ee
The parameters $\Lambda$ and $\beta$ in turn are defined by two
additional equations
\be{4sadd1neg}
i(\Lambda-\lambda_0)+\frac{1}{2\pi}\frac{\partial}{\partial\beta}
\int_{-\infty}^\infty (\lambda_0-\lambda)
\sign(\Lambda-\lambda)
\ln\left[\frac{e^{\frac{\varepsilon(\lambda)}{T}}-
e^{w(\lambda)\sign(\lambda-\Lambda)}}
{e^{\frac{\varepsilon(\lambda)}{T}}+1}\right]
\,d\lambda=0,
\ee
\be{4sadd2neg}
i(\beta+\Lambda-\lambda_0)+\frac{1}{2\pi}
\frac{\partial}{\partial\Lambda}
\int_{-\infty}^\infty (\lambda_0-\lambda)
\sign(\Lambda-\lambda)
\ln\left[\frac{e^{\frac{\varepsilon(\lambda)}{T}}-
e^{w(\lambda)\sign(\lambda-\Lambda)}}
{e^{\frac{\varepsilon(\lambda)}{T}}+1}\right]
\,d\lambda=0.
\ee
We also took into account that $x=2t\lambda_0$.

For positive chemical potential formul\ae\ are rather similar, but
there is a set of differences.  The correlation radius is given by
\ba{4radpos}
&&{\dis \hspace{-2cm}
-t/r_+=2it(\Lambda-\lambda_0)\beta+
\frac{it}2(\Lambda_1^2+\Lambda_2^2-2h)
-\frac{ix}2(\Lambda_1+\Lambda_2)}\non
&&{\dis
+\frac{1}{2\pi}\int_{-\infty}^\infty(x-2\lambda t)
\sign(\Lambda-\lambda)
\ln\left[\frac{e^{\frac{\varepsilon(\lambda)}{T}}-
e^{w(\lambda)\sign(\lambda-\Lambda)}}
{e^{\frac{\varepsilon(\lambda)}{T}}+1}\omega(\lambda)\right]
\,d\lambda.}
\ea
Hereby the function $w(u)$ satisfies the integral equation
\ba{4inteqpos}
&&{\dis\hspace{-2cm}
w(u)=\frac12(\xi(\Lambda_1,u)+\xi(\Lambda_2,u))
+i\beta K(\Lambda,u)}\non
&&{\dis
-\frac{1}{2\pi}\int_{-\infty}^\infty K(u,\lambda)
\sign(\Lambda-\lambda)
\ln\left[\frac{e^{\frac{\varepsilon(\lambda)}{T}}-
e^{w(\lambda)\sign(\lambda-\Lambda)}}
{e^{\frac{\varepsilon(\lambda)}{T}}+1}\omega(\lambda)
\right]
\,d\lambda,}
\ea
where $\Lambda_1$ and $\Lambda_2$ are the roots of the
equation \eq{4classeq}
\be{4roots}
\varepsilon(\Lambda_i)=T\sign(\Lambda-\Lambda_i)
w(\Lambda_i),\qquad
\Lambda_1\to-q_T,\quad\Lambda_2\to q_T,
\qquad\mbox{at}\qquad T\to0,
\ee
and
$$
\omega(\lambda)=\sign(\lambda-\Lambda_1)
\sign(\lambda-\Lambda_2).
$$
The parameters $\Lambda$ and $\beta$ can be found from additional
equations
\be{4sadd1pos}
i(\Lambda-\lambda_0)+\frac{1}{2\pi}\frac{\partial}{\partial\beta}
\int_{-\infty}^\infty (\lambda_0-\lambda)
\sign(\Lambda-\lambda)
\ln\left[\frac{e^{\frac{\varepsilon(\lambda)}{T}}-
e^{w(\lambda)\sign(\lambda-\Lambda)}}
{e^{\frac{\varepsilon(\lambda)}{T}}+1}\omega(\lambda)\right]
\,d\lambda=0,
\ee
\be{4sadd2pos}
i\beta+\frac{1}{2\pi}
\frac{\partial}{\partial\Lambda}
\int_{-\infty}^\infty (\lambda_0-\lambda)
\sign(\Lambda-\lambda)
\ln\left[\frac{e^{\frac{\varepsilon(\lambda)}{T}}-
e^{w(\lambda)\sign(\lambda-\Lambda)}}
{e^{\frac{\varepsilon(\lambda)}{T}}+1}\omega(\lambda)\right]
\,d\lambda=0.
\ee
Thus, the exponential decay of the correlation function
\eq{0tempcorrel} is defined by the function $w(u)$. This function can
be found from the system of integral equations, which structure is
rather similar to the Yang--Yang equation \eq{0YY}. Therefore it is
quite possible that $w(u)$ can be expressed in terms of observables
of the one-dimensional Bose gas. In the next section we shall
demonstrate this for some limiting cases.

In the conclusion of this section we consider free fermionic limit:
$c\to\infty$. Then $K(\lambda,\mu)=\xi(\lambda,\mu)=0$
(see \eq{4xiK}), and hence, $w(u)=0$. Besides
$\varepsilon(\lambda)=\lambda^2-h$, $\Lambda=\lambda_0$, $\beta=0$
and $\Lambda_{1,2}=\mp\sqrt h$. The integral term in the equations
\eq{4radneg}, \eq{4radpos} turns into
\be{4ff}
\frac{1}{2\pi}\int_{-\infty}^\infty|x-2\lambda t|
\ln\left|\frac{e^{\frac{\varepsilon(\lambda)}{T}}-1}
{e^{\frac{\varepsilon(\lambda)}{T}}+1}\right|
\,d\lambda.
\ee
In the case of positive chemical potential this expression completely
describes the asymptotic behavior of the correlation function, since
$\Lambda_1^2+\Lambda_2^2-2h=0$. For $h<0$ we have additional
oscillating term $-it(\lambda_0^2+h)$. These formul\ae\ exactly
reproduce the results of \cite{IIKV}.

However the free fermionic limit is not too representative.
Indeed, in this case all the commutation relations between operators
$p$ and $q$ \eq{4commut} become trivial, therefore we could
put from the very beginning all the dual fields equal to zero
in the equations \eq{4main1}, \eq{4main2}.

\section{Limiting cases}

In this section we consider some limiting cases of the correlation
function \eq{0tempcorrel}. The first case corresponds to the
autocorrelation. In spite of we considered the limit $x\to\infty$,
$t\to\infty$, in fact all the asymptotic analysis performed in
\cite{IS} was based on the assumption that the $x/2t=\lambda_0$ is
finite. In particular $\lambda_0$ may be equal to zero, what describes
the case $x=0$.

Consider, for instance, the case $h<0$. It is easy to check that for
$\lambda_0=0$ one can put $\Lambda=\beta=0$ in the equations
\eq{4inteqneg}--\eq{4sadd2neg}. Indeed, it follows from
\eq{4inteqneg} that for zero values of these parameters
$w(u)$ is an odd function. Differentiating
\eq{4inteqneg} with respect to $\Lambda$ and $\beta$ at zero point we
find that the derivatives $\partial w/\partial\Lambda$ and $\partial
w/\partial\beta$ are even functions of $u$. Then the equations
\eq{4sadd1neg}, \eq{4sadd2neg} are valid automatically, and the result
slightly simplifies
\be{5avtoneg}
r_-^{-1}=ih-\frac{2}{\pi}\int_0^\infty
\lambda\ln\left[
\frac{e^{\frac{\varepsilon(\lambda)}{T}}
-e^{w(\lambda)}}{e^{\frac{\varepsilon(\lambda)}{T}}+1}
\right]\,d\lambda,
\ee
where $w(u)$ is the solution of the equation
\be{5avtoeqneg}
w(u)=-\xi(u,0)+\frac{1}{2\pi}\int_0^\infty
\Bigl(K(\lambda,u)-K(\lambda,-u)\Bigr)\ln\left[
\frac{e^{\frac{\varepsilon(\lambda)}{T}}
-e^{w(\lambda)}}{e^{\frac{\varepsilon(\lambda)}{T}}+1}
\right]\,d\lambda.
\ee

Similarly for positive chemical potential we obtain
\be{5avtopos}
r_+^{-1}=i(h-\Lambda_1^2)-\frac{2}{\pi}\int_0^\infty
\lambda\ln\left[
\frac{e^{\frac{\varepsilon(\lambda)}{T}}
-e^{w(\lambda)}}{e^{\frac{\varepsilon(\lambda)}{T}}+1}
\omega(\lambda)\right]\,d\lambda,
\ee
and
\be{5avtoeqpos}
w(u)=\frac{1}{2}(\xi(\Lambda_1,u)-\xi(\Lambda_1,-u))
+\frac{1}{2\pi}\int_0^\infty
\Bigl(K(\lambda,u)-K(\lambda,-u)\Bigr)\ln\left[
\frac{e^{\frac{\varepsilon(\lambda)}{T}}
-e^{w(\lambda)}}{e^{\frac{\varepsilon(\lambda)}{T}}+1}
\omega(\lambda)\right]\,d\lambda.
\ee
Hereby it turns out that $\Lambda_2=-\Lambda_1$.

The most interesting is the low temperature limit for the case
$h>0$. In this limit we have a possibility to compare our results with
the known ones. Indeed, as we have mentioned already, at
$T=0$ the asymptotics of the correlation function must be power-like.
This means that the correlation radius $r_+\to\infty$.
Besides, the correlation radius was computed for small temperature in
the linear approximation in \cite{BPZ}. Thus, one can consider the
low temperature limit as a good test for our results. Besides we have
a possibility to make more precise estimates of the results obtained
previously.

First, let us present the system of equations, describing the ground
state for $h>0$ \cite{LL}. Recall the Yang--Yang equation \eq{0YY}
\be{5YY}
\varepsilon(\lambda)=\lambda^2-h-\frac{T}{2\pi}
\int_{-\infty}^{\infty}
K(\lambda,\mu)\ln
\left(1+e^{-\frac{\varepsilon(\mu)}{T}}
\right)d\mu.
\ee
The function $\varepsilon(\lambda)$ has two real roots
$\varepsilon(\pm q_T)=0$,  and
$\varepsilon(\lambda)>0$, if $|\lambda|>q_T$, and
$\varepsilon(\lambda)<0$ if  $|\lambda|<q_T$. At $T\to0$
the roots $\pm q_T$ tend to some fixed values (corresponding to the
boundaries of the Fermi sphere): $q_T\to q$. It is easy to see that
$$
\exp\left\{-\frac{\varepsilon(\mu)}{T}\right\}
\stackrel{T\to0}{\longrightarrow}\left\{
\begin{array}{ll}
0,&\qquad|\mu|>q,\\
\infty,&\qquad|\mu|<q.
\end{array}\right.
$$
Thus the equation \eq{5YY0} turns into
\be{5YY0}
\varepsilon_0(\lambda)=\lambda^2-h+\frac{1}{2\pi}
\int_{-q}^q K(\lambda,\mu)\varepsilon_0(\mu)\,d\mu,
\qquad \varepsilon_0(\pm q)=0.
\ee
Here and further we denote observables at zero temperature by
the subscript $0$. Similarly the equations for the density and the
scattering phase have the form
\be{5rho0}
\rho_0(\lambda)=\frac{1}{2\pi}+\frac{1}{2\pi}\int_{-q}^q
K(\lambda,\mu)\rho_0(\mu)\,d\mu,
\ee
\be{5phase0}
F_0(\lambda,\nu)=\frac{1}{2\pi}\int_{-q}^q
K(\lambda,\mu)F_0(\mu,\nu)\,d\mu+
\frac{1}{2\pi i}\xi(\lambda,\nu).
\ee
The derivative of the momentum and the velocity as before are equal to
\be{5mement0}
\frac{\partial k_0(\lambda)}{\partial\lambda}=
2\pi \rho_0(\lambda),\qquad
v_0(\lambda)=\frac{\varepsilon'_0(\lambda)}{k'_0(\lambda)}.
\ee
Let us give also the equation for the resolvent of the integral
operator, entering the equations \eq{5YY0}--\eq{5phase0}:
$(I+R)(I-\frac{1}{2\pi}K)=I$
\be{5res0}
R(\lambda,\nu)=\frac{1}{2\pi}\int_{-q}^q
K(\lambda,\mu)R(\mu,\nu)\,d\mu+
\frac{1}{2\pi}K(\lambda,\nu).
\ee
It is easy to see that $R(\lambda,\mu)=-\partial_\mu
F_0(\lambda,\mu)$.  Besides, comparing the equations \eq{5rho0} and
\eq{5phase0}  and using $K(\lambda,\mu)=-i\partial_\lambda
\xi(\lambda,\mu)$, we find:
\be{5rhophase}
2\pi\rho_0(\lambda)=1+F_0(\lambda,-q)-F_0(\lambda,q).
\ee

Similarly to the Yang--Yang equation the integral equation for the
function $w(u)$ becomes linear in the limit $T\to0$.
Let, for example, $\Lambda_1<\Lambda_2<\Lambda$. Then instead of
\eq{4inteqpos} we have at $T=0$
\be{5eqT0}
w_0(u)=\frac{1}{2}(\xi(-q,u)+\xi(q,u))+i\beta K(u,\Lambda)
+\frac{1}{2\pi}\int_{-q}^q K(u,\lambda)w_0(\lambda)\,d\lambda.
\ee
Here we have used that $\Lambda_2=-\Lambda_1=q$ for $T=0$. Comparing
the equation \eq{5eqT0} with \eq{5phase0} and \eq{5res0}, we find
\be{5solT0}
w_0(u)=-i\pi(F_0(u,-q)+F_0(u,q))+2\pi i\beta R(u,\Lambda).
\ee
The restrictions for the numbers $\Lambda$ and $\beta$ have the
following form:
\be{5dopbeta}
i(\Lambda-\lambda_0)-\frac{1}{2\pi}
\frac{\partial}{\partial\beta}\int_{-q}^q
(\lambda_0-\lambda)w_0(\lambda)\,d\lambda=0,
\ee
\be{5dopLambda}
i\beta-\frac{1}{2\pi}
\frac{\partial}{\partial\Lambda}\int_{-q}^q
(\lambda_0-\lambda)w_0(\lambda)\,d\lambda=0.
\ee
It follows immediately from \eq{5solT0} and \eq{5dopLambda} that
$\beta=0$, and hence,
\be{5solT01}
w_0(u)=-i\pi(F_0(u,-q)+F_0(u,q))
\ee
Substituting this into \eq{4radpos} (where the logarithm also turns
into linear function), we find
\be{5radpos}
-t/r_+=it(q^2-h)+\frac{i}{2}\int_{-q}^q(x-2\lambda t)
(F_0(\lambda,q)+F_0(\lambda,-q))\,d\lambda.
\ee
It is easy to see that the r.h.s. of \eq{5radpos} is equal to zero
identically. Indeed, due to \eq{5phase0} the integral in the r.h.s.
of \eq{5radpos} can be represented in the form
$$
\frac{1}{4\pi}\int_{-q}^q(x-2\lambda t)
\bigl(\delta(\lambda-\mu)+R(\lambda,\mu)\bigr)
(\xi(\mu,q)+\xi(\mu,-q))\,d\lambda d\mu.
$$
Acting by the resolvent to the left and using \eq{5YY0}--\eq{5mement0},
we obtain
$$
\frac{1}{4\pi}\int_{-q}^q(xk'_0(\lambda)-t\varepsilon'_0(\lambda))
(\xi(\lambda,q)+\xi(\lambda,-q))\,d\lambda.
$$
Since $k'_0(\lambda)$ is an even function, then the coefficient at
$x$ is equal to zero. The remaining integral  after the integration
by parts gives
$$
\frac{it}{4\pi}\int_{-q}^q\varepsilon_0(\lambda)
(K(\lambda,q)+K(\lambda,-q))\,d\lambda.
$$
Now one can use the equation \eq{5YY0}  and the condition
$\varepsilon_0(\pm q)=0$, after that we obtain
\be{5radT0}
r_+^{-1}=0, \qquad\mbox{for}\qquad T=0.
\ee
Thus our formul\ae\ do give the correct result at zero temperature.

Recall that the introducing of the dual fields is equivalent in some
sense to factorization of the $S$-matrix. It is natural to expect that
under the averaging of the dual fields the objects have to appear, having
direct relation to the scattering matrix. However, while the
factorization takes place at the bare Fock vacuum, the averaging
is performed already in the thermodynamic limit, i.e. at the
physical vacuum. It is remarkable that at this rate instead of the
`bare' $S$-matrix \eq{2Smatr} the function $w_0(u)$ arises, which
can be expressed in terms of the `dressed' scattering phase
\eq{5phase0} at the boundary of the Fermi sphere.

Observe also, that in the calculations presented above, we did not
need the explicit value of $\Lambda$. However, it is not difficult
to find it. It follows from the equation \eq{5dopbeta} that
$$
\Lambda-\lambda_0-\int_{-q}^q(\lambda_0-\lambda)
R(\lambda,\Lambda)\,d\lambda=0,
$$
and we immediately find
\be{5phys}
\lambda_0k'_0(\Lambda)-\frac{1}{2}\varepsilon'_0(\Lambda)=0,
\qquad\mbox{or}\qquad v_0(\Lambda)=\frac{x}{t}.
\ee
Thus the quantity $\Lambda$ has a pure physical sense: a particle
possessing velocity $v_0(\Lambda)$, passes the distance
$x$ spending the time $t$.

In order to compute the correlation radius for small, but not zero
temperature, more accurate estimates of the integrals containing
logarithmic function are necessary. The example of such calculations
is given in the Appendix A. It is clear, however, that all low
temperature corrections to the correlation radius, as well as to
the integral equation for the function $w(u)$, are provided by
vicinities of the points $\pm q$, defining the boundary of the Fermi
sphere. Let us present here the integral equation for the function
$w(u)$, arising in the first order with respect to $T$. We shall look
for the function $w(u)$ as a series
$$
w(u)=w_0(u)+Tw_1(u)+T^2w_2(u)+\dots
$$
Here $w_0(u)$ is given by  \eq{5solT01}. For
$w_1(u)$ we have
\be{5eqw1}
w_1(u)-\frac{1}{2\pi}\int_{-q}^qK(u,\lambda)w_1(\lambda)
\,d\lambda=
-\frac{K(u,q)}{4\pi\varepsilon'_0(q)}(w_0(q)+i\pi)^2
-\frac{K(u,-q)}{4\pi\varepsilon'_0(q)}(w_0(-q)-i\pi)^2.
\ee
Hence
\be{5solw1}
w_1(u)=\frac{\pi^2}{2\varepsilon'_0(q)}
\bigl(1-F_0(q,q)-F_0(q,-q)\bigr)^2(R(u,q)+R(u,-q)).
\ee
Here we took into account, first, that
$F_0(-\lambda,-\mu)=-F_0(\lambda,\mu)$.  Second, we have put already the
parameter $\beta=0$. On can easily check that it is always so in the
framework of the low temperature approximation. Indeed, for any order
of $T$ the dependency of $w(u)$ on the sign-functions
$\sign(\Lambda-u)$ disappears. The parameters $\Lambda$ and $\beta$
enter the equation for $w(u)$ only in the combination  $\beta
K(\Lambda,u)$ (see \eq{Peq}). Therefore the derivative $\partial
w/\partial\Lambda$ always turns out to be proportional to $\beta$,
and hence, the equation \eq{4sadd2pos} is valid for $\beta=0$.

Substituting $w_1(u)$ into the answer for the correlation radius, after
simple algebra we arrive at
\be{5radw1}
r_+^{-1}=\frac{\pi T}{4v_0}\bigl(1-F_0(q,q)-F_0(q,-q)\bigr)^2
\bigl(|x-v_0t|+|x+v_0t|\bigr).
\ee
Here $v_0=v_0(q)$. This result is given for arbitrary location
of the roots $\Lambda_1$ and $\Lambda_2$ with respect to the point
$\Lambda$. In the case $\Lambda_1<\Lambda_2<\Lambda$, which we
considered from the very beginning, one can remove the symbols of the
absolute value.

Comparing \eq{5radw1} with the result obtained in \cite{BPZ}
\be{5radBPZ}
r_+^{-1}=\frac{\pi T}{4v_0}\cdot
\frac{|x-v_0t|+|x+v_0t|}{(2\pi\rho_0(q))^2},
\ee
we obtain an identity
\be{5ident1}
\frac{1}{2\pi\rho_0(q)}= 1-F_0(q,q)-F_0(q,-q).
\ee
At the first sight such the non-linear relationship between the
solutions of the linear integral equations looks rather strange.
However, taking into account a non-linear identity for the scattering
phase \cite{KS3}, \cite{S2}
\be{5ident2}
\bigl(1-F_0(q,q)\bigr)^2-F_0^2(q,-q)=1
\ee
and the equation \eq{5rhophase}, we make sure of the validity of the
identity \eq{5ident1}. Thus, in the linear approximation with respect
to the temperature our result for the correlation radius exactly
coincides with the result obtained in the framework of the conformal
field theory.

The computation of higher temperature corrections does not meet
serious difficulties. In the all orders $T^n$ the equations for
$w_n(u)$ remain linear (see \eq{Peq}), and their solutions can be
written explicitly in terms of the resolvent of the operator
$I-\frac{1}{2\pi}K$ and its derivatives. Respectively the correlation
radius is expressed in terms of functions $k'_0(q)$,
$\varepsilon'_0(q)$, $F_0(q,\pm q)$ and their derivatives. In
particular the correction to $r_+^{-1}$ of the order $T^2$ vanishes.
Thus the result \eq{5radw1} is valid up to the terms of order
$T^3$.

\section*{Conclusion}

We have considered the large time and long distance asymptotic
behavior of the correlation function of the quantum one-dimensional
Bose gas. As we have shown, the asymptotics can be expressed in terms
of the integral equation solutions, which are closely related to the
equations of the thermodynamic Bethe Ansatz. In the low temperature
limit we were able even to describe the solutions of mentioned
equations in terms of observables of the model. We would like to draw
attention of the reader to the fact, that these integral equations arise
under the averaging of the dual fields. While in the earlier works these
quantum operators played purely auxiliary role and mostly were used for
solving of certain combinatorial problems, the results obtained now
allows us to have a look at these objects from a new point of view.
In particular, it is hardly accidentally that at the zero temperature
the solutions of the integral equations depend on the scattering
phase. Apparently the role of the dual fields is much more
fundamental, than it was understood before.

We would like to thank A.~R.~Its and V.~E.~Korepin for useful
discussions.

\appendix
\section{Appendix}

As an example of the low temperature expansion we consider the
integral, entering the Yang--Yang equation
\be{Pint}
I=T\int_{-\infty}^\infty K(\lambda,\mu)
\ln\left(1+e^{-\frac{\varepsilon(\mu)}T}\right)\,d\mu.
\ee
Obviously, at $T=0$  only the interval $[-q,q]$ contributes into
this integral
\be{PintT0}
I_{T=0}=-\int_{-q}^q K(\lambda,\mu)
\varepsilon(\mu)\,d\mu,  \qquad
\varepsilon(\pm q)=0\qquad\mbox{при}\qquad T=0.
\ee
At small, but finite temperature the corrections to this expression
arise due to the contributions of vicinities of  the points $\pm q$.
Consider, for instance, the integral
\be{Pintplus}
I_+=T\int_{q_T}^\infty K(\lambda,\mu)
\ln\left(1+e^{-\frac{\varepsilon(\mu)}T}\right)\,d\mu,
\ee
where $\varepsilon(q_T)=0$. Making the replacement of
variables $\varepsilon(\mu)=Tz$, we obtain
\be{Pinta}
I_+=T^2\int_{0}^\infty \varphi(\lambda,Tz)
\ln\left(1+e^{-z}\right)\,dz,
\ee
where $\varphi(\lambda,Tz)=K(\lambda,\mu)/\varepsilon'(\mu)$. The
integral \eq{Pinta} can be expanded now into the series with respect
to $T$. We have
\be{Pintb}
I_+=\sum_{n=0}^{\infty} T^{n+2}\varphi^{(n)}(\lambda,0)
(1-2^{-n-1})\zeta(n+2),
\ee
or in terms of the functions $K(\lambda,\mu)$ and $\varepsilon(\mu)$
\be{Pintc}
I_+=\sum_{n=0}^{\infty} T^{n+2}(1-2^{-n-1})\zeta(n+2)
D^n_{q_T}\frac{K(\lambda,q_T)}{\varepsilon'(q_T)},
\qquad\mbox{where}\qquad D_\lambda=\frac{1}{\varepsilon'(\lambda)}
\frac{\partial}{\partial\lambda}.
\ee
Similarly all the remaining contributions from vicinities of the points
$\pm q_T$ can be computed. Then one need to substitute the expansion
obtained into the Yang--Yang equation and to put
\be{Pdec}
\begin{array}{l}
{\dis
\varepsilon(\lambda)=
\sum_{n=0}^{\infty}T^n\varepsilon_n(\lambda),}\non
{\dis
q_T=q+\sum_{n=1}^{\infty}T^n q_n.}
\end{array}
\ee
The arising infinite set of equations can be solved via recursion.
As a result the functions $\varepsilon_n$ and parameters $q_n$
are expressed in terms of $\varepsilon'_0(q)$  and the resolvent
$R(\lambda,\pm q)$.

Similarly all the integrals of the section 4 can be decomposed into
series. Let us present the complete expansion with respect to
the temperature of the integral equation \eq{4inteqpos}
for the function $w(u)$
\ba{Peq}
&&{\dis\hspace{-1cm}
w(u)-\frac{1}{2\pi}\int_{-q}^q K(u,\lambda)w(\lambda)
\,d\lambda
=\frac{1}{2}\Bigl(\xi(-q,u)+\xi(q,u)\Bigr)
+i\beta K(\Lambda,u)
}\non
&&{\dis
-\frac{1}{2\pi}\sum_{\gamma=\pm1}\sum_{n=2}^{\infty}
\frac{(-2\pi i\gamma)^nT^{n-1}}{n!}D^{n-2}_{q_T}
\frac{K(u,\gamma q_T}{\varepsilon'(q_T)}\left[
B_n\left(1+\gamma\frac{w(\gamma q_T)}{2\pi i}\right)-
B_n(1/2)\right].}
\ea
Here $B_n(z)$ are the Bernulli polynomials. It is yet necessary to
substitute into this equation decompositions \eq{Pdec}, after that
functions $w_n(u)$  can be evaluated in terms of
$\varepsilon'_0(q)$, $R(\lambda,\pm q)$ and their derivatives.

\end{document}